\documentclass[a4paper,twocolumn,superscriptaddress,pre, aps, floatfix ]{revtex4-1}
\usepackage{amssymb}
\usepackage{amsmath}
\usepackage{marvosym}
\usepackage{graphicx}
\usepackage{color}
\usepackage{url}
\usepackage{lipsum}
\usepackage{soul}
\usepackage{multirow}
\usepackage[colorlinks=true,allcolors=blue]{hyperref}

\begin{document}

\title{Feature-enriched hyperbolic network geometry}

\author{Roya Aliakbarisani}
\email{roya\_aliakbarisani@ub.edu}
\affiliation{Departament de F\'isica de la Mat\`eria Condensada, Universitat de Barcelona, Mart\'i i Franqu\`es 1, E-08028 Barcelona, Spain}
\affiliation{Universitat de Barcelona Institute of Complex Systems (UBICS), Barcelona, Spain}

\author{M. \'Angeles Serrano}
\email{marian.serrano@ub.edu}
\affiliation{Departament de F\'isica de la Mat\`eria Condensada, Universitat de Barcelona, Mart\'i i Franqu\`es 1, E-08028 Barcelona, Spain}
\affiliation{Universitat de Barcelona Institute of Complex Systems (UBICS), Barcelona, Spain}
\affiliation{Instituci\'o Catalana de Recerca i Estudis Avan\c{c}ats (ICREA), Passeig Llu\'is Companys 23, E-08010 Barcelona, Spain}

\author{Mari\'an Bogu\~n\'a}
\email{marian.boguna@ub.edu}
\affiliation{Departament de F\'isica de la Mat\`eria Condensada, Universitat de Barcelona, Mart\'i i Franqu\`es 1, E-08028 Barcelona, Spain}
\affiliation{Universitat de Barcelona Institute of Complex Systems (UBICS), Barcelona, Spain}

\begin{abstract}
Graph-structured data provide a comprehensive description of complex systems, encompassing not only the interactions among nodes but also the intrinsic features that characterize these nodes. These features play a fundamental role in the formation of links within the network, making them valuable for extracting meaningful topological information. Notably, features are at the core of deep learning techniques such as Graph Convolutional Neural Networks (GCNs) and offer great utility in tasks like node classification, link prediction, and graph clustering. In this paper, we present a comprehensive framework that treats features as tangible entities and establishes a bipartite graph connecting nodes and features. By assuming that nodes sharing similarities should also share features, we introduce a hyperbolic geometric space where both nodes and features coexist, shaping the structure of both the node network and the bipartite network of nodes and features. Through this framework, we can identify correlations between nodes and features in real data and generate synthetic datasets that mimic the topological properties of their connectivity patterns. The approach provides insights into the inner workings of GCNs by revealing the intricate structure of the data.
\end{abstract}

\maketitle
\section{Introduction}

The nature of link formation in complex networks has been a recurrent theme during the last two decades of research in network science. Understanding the key factors contributing to the emergence of interactions among individual elements is the first step to understanding the system as a whole and, thus, the emerging behaviors that arise from such interactions. Beyond purely topological link formation mechanisms, such as preferential attachment~\cite{barabasi1999emergence}, nodes in a network have well-defined features that also play a role during the link formation process. In this context, network geometry~\cite{Boguna2021} offers a simple yet powerful approach to explaining the topology of networks in terms of underlying metric spaces that effectively encode topological properties and intrinsic node attributes~\cite{serrano2008similarity,krioukov2010hyperbolic,papadopoulos2012popularity}. Only recently, the explosion of graph-structured data (networks with annotated information) is being used to understand the emergence of communities in networks~\cite{Newman2016, Peel2017, Bassolas2022, Mucha2019, Smith2017} or their percolation properties~\cite{Artime2021}.

Graph-structured data is particularly relevant for deep learning techniques. Specifically, Graph Convolutional Neural Networks (GCNs) have emerged as a powerful tool for effectively modeling and analyzing graph data, enabling us to leverage the expressive power of deep learning on irregular and non-Euclidean domains~\cite{zhou2020, wu2021}. GCNs are an extension of classical Convolutional Neural Networks (CNNs) that are designed to work with graph-structured data. While CNNs are effective at extracting spatial patterns from grid-like data, GCNs go beyond by considering the graph structure. GCNs aggregate information from the neighborhood of each node in a graph, allowing them to propagate information and capture the graph topology. This makes GCNs particularly useful for tasks like node classification, link prediction, graph clustering, or recommendation systems, to name just a few.

Despite their undeniable effectiveness, machine learning techniques, in particular CNNs  and GCNs, are criticized for their lack of explainability, a problem referred to as the black box problem~\cite{castelvecchi2016can}. An implicit assumption made by GCNs is that there must exist correlations between connected (or topologically close) nodes in the graph so that they are ``similar", and similar nodes should share common features. Only when this is the case, GCNs are able to detect patterns in the data. Thus, to solve the black box problem, we must first understand in detail the structure of the data that feeds GCNs. 

In this paper, we introduce a simple yet comprehensive framework to describe real graph-structured datasets. Our approach has two critical contributions. First, we consider features as real entities that define a bipartite graph of nodes connected to features which, for real systems, shows a complex topological organization. Second, we assume that if two nodes are similar when they share features, then two features are also similar if they share nodes. Following this reasoning, we introduce a geometric similarity space where both nodes and features coexist, shaping the structure of both the network between nodes and the bipartite network of nodes and features. Using this framework, we are able to detect correlations between nodes and features in real data and generate synthetic datasets with the same topological properties.

\section{Results}

A typical graph-structured dataset consists of a set of $N_n$ nodes forming a complex network $\mathcal{G}_n$ and a set of $N_f$ features associated with the same set of nodes. The features are usually binarized, so the set of features for a given node $i$ is represented as a vector $\vec{f}_i \in \mathbb{R}^{N_f}$ with entries of zero or one, indicating the presence or absence of a particular feature. For example, the Cora dataset is a standard benchmark used in GCN studies. It is defined by a citation network among scientific publications --or nodes-- and each publication is characterized by a vector, where the entries indicate the presence or absence of specific words --or features-- from a unique dictionary.

To fully characterize such complex graph-structured data, we must first understand the complex network $\mathcal{G}_n$ that defines the relationships between nodes. In CNNs applied to images, for instance, this network is defined by the nearest neighbors in a two-dimensional grid of pixels. However, in complex graph-structured data, the relationships between nodes are better described by a complex network with intricate topological properties. Our research over the last decade has shown that complex networks, such as the ones of interest in this context, can be accurately characterized using geometric random graph models~\cite{Boguna2021}. In these models, nodes are positioned in a metric space, and the probability of connection between nodes depends on their distances in this space. This approach has led to the emergence of network geometry as a field, providing a comprehensive understanding of real complex networks. Geometric models in a latent hyperbolic metric space have proven effective in generating networks with realistic topological properties, including heterogeneous degree distributions~\cite{serrano2008similarity,krioukov2010hyperbolic,gugelmann2012random}, clustering~\cite{krioukov2010hyperbolic,gugelmann2012random,candellero2016clustering,Fountoulakis2021}, small-worldness~\cite{abdullah2017typical,friedrich2018diameter,muller2019diameter}, percolation~\cite{serrano2011percolation,fountoulakis2018law}, spectral properties~\cite{kiwi2018spectral}, and self-similarity~\cite{serrano2008similarity}. They have also been extended to encompass growing networks~\cite{papadopoulos2012popularity}, weighted networks~\cite{allard2017geometric}, multilayer networks~\cite{kleineberg2016hidden,Kleineberg2017}, networks with community structure~\cite{zuev2015emergence,garcia-perez:2018aa,muscoloni2018nonuniform}, and serve as the basis for defining a renormalization group for complex networks~\cite{garcia-perez2018multiscale,Zheng:2021aa}. In this case, this approach is particularly interesting as it naturally introduces the concept of an underlying similarity space, allowing the unambiguous quantification of similarity between nodes.

\begin{figure}[t]
	\centering
	\includegraphics[width=\columnwidth]{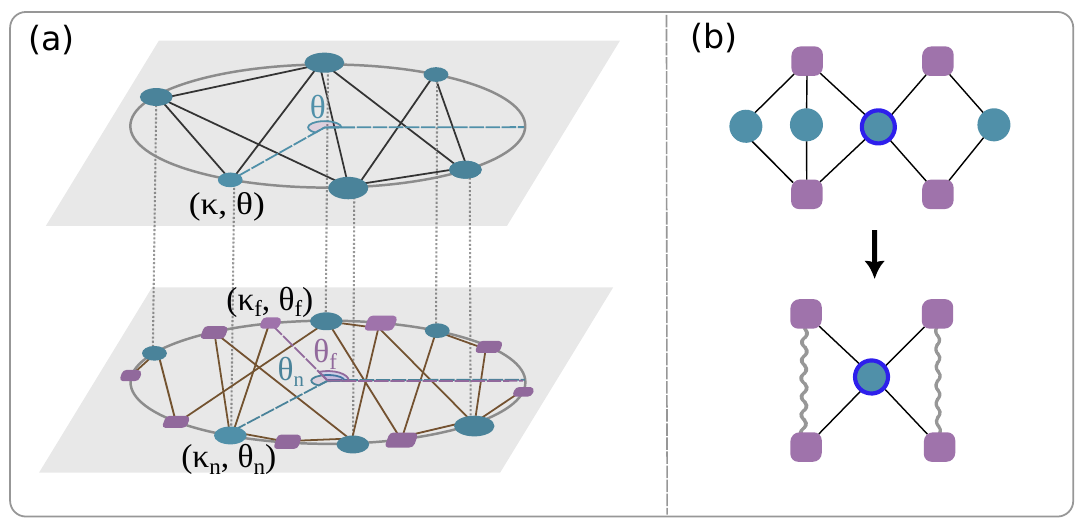}
	\caption{{\bf Illustration of the bipartite model used to describe graph structured data.} The sketch at the top of panel (a) depicts the generation of a network using the $\mathbb{S}^1$ model. The bottom part illustrates the generation of the bipartite network between nodes (green circles), keeping the same angular coordinates, and features (rounded purple squares). Panel (b) illustrates the method for measuring the bipartite clustering. For example, the node at the center is connected to four different features. Two features are considered connected if they share at least a common node other than the central node. The bipartite clustering of the node is then calculated by determining the fraction of connected pairs of features, following the standard definition of clustering coefficient in unipartite networks. The same definition applies to the bipartite clustering coefficient of features.}
	\label{fig:1}
\end{figure}
\subsubsection{Modeling the network. The $\mathbb{S}^1/\mathbb{H}^2$ model}

To describe the network between nodes $\mathcal{G}_n$, we employ the $\mathbb{S}^1$ model, also known as the geometric soft configuration model~\cite{serrano2008similarity,krioukov2010hyperbolic,Boguna:2020fj,serrano_boguna_2022}. In this model, each node is assigned two hidden variables $(\kappa, \theta)$ that determine its expected degree and position on a one-dimensional sphere of radius $R = N_n/2\pi$. This sphere represents the abstraction of the similarity space where nodes are placed~\footnote{The model can also be defined on spheres of arbitrary dimensions. However, the one-dimensional case captures the most relevant topological properties}. The connection probability between two nodes with hidden variables $(\kappa, \theta)$ and $(\kappa', \theta')$ is defined as follows:
\begin{equation}
    p(\kappa,\kappa',\Delta \theta)=\frac{1}{1+\chi^\beta} \;  \; \mbox{with}  \; \; \chi \equiv \frac{R \Delta \theta}{\mu \kappa \kappa'},
    \label{pkkprima}
\end{equation}
where $\Delta \theta = \pi - |\pi - |\theta - \theta'||$ represents the angular separation between the nodes, $\beta > 1$~\footnote{The case $\beta < 1$ has been thoroughly analyzed in~\cite{Jasper_2022}} is the inverse of the temperature of the graph ensemble and determines the level of clustering in the network, and $\mu = \frac{\beta}{2\pi \langle k \rangle} \sin{\frac{\pi}{\beta}}$ is a parameter that fixes the average degree $\langle k \rangle$ (see the top of panel (a) in Fig.~\ref{fig:1}). The hidden variables of the nodes can either be generated from an arbitrary probability density $\rho(\kappa, \theta)$ if the goal is to create synthetic networks or can be inferred from a real network by maximizing the likelihood of the model to reproduce the desired real network. In this work, we use the latter approach through the embedding tool called Mercator~\cite{GarciaPerez2019}.

Interestingly, the $\mathbb{S}^1$ model is isomorphic to a purely geometric model in the hyperbolic plane, the $\mathbb{H}^2$ model~\cite{krioukov2010hyperbolic}. By mapping the expected degree of a given node to a radial coordinate as
\begin{equation}
r=R_{\mathbb{H}^2}-2\ln{\frac{\kappa}{\kappa_0}}\; \; \mbox{with} \; \; R_{\mathbb{H}^2}=2\ln{\frac{2R}{\mu \kappa_0^2}}
\end{equation}
and $\kappa \ge \kappa_0$, the connection probability Eq.~\eqref{pkkprima} becomes
\begin{equation}
p(x)=\frac{1}{1+e^{\frac{\beta}{2}(x-R_{\mathbb{H}^2})}} \; \; \mbox{with} \; \; x=r+r'+2\ln{\frac{\Delta \theta}{2}} 
\label{eq:pkkhyperbolic}
\end{equation}
and where $x$ is a very good approximation of the hyperbolic distance between two points at radial coordinates $r$ and $r'$ and separated by an angular distance $\Delta \theta$. Thanks to this equivalence, we can use either version of the model depending on the particular application. The $\mathbb{S}^1$ model is more convenient for performing analytic calculations whereas the $\mathbb{H}^2$ model is more suited for tasks like greedy routing~\cite{boguna2010sustaining} and for visualization purposes.

\begin{figure}[t]
	\centering
	\includegraphics[width=\columnwidth]{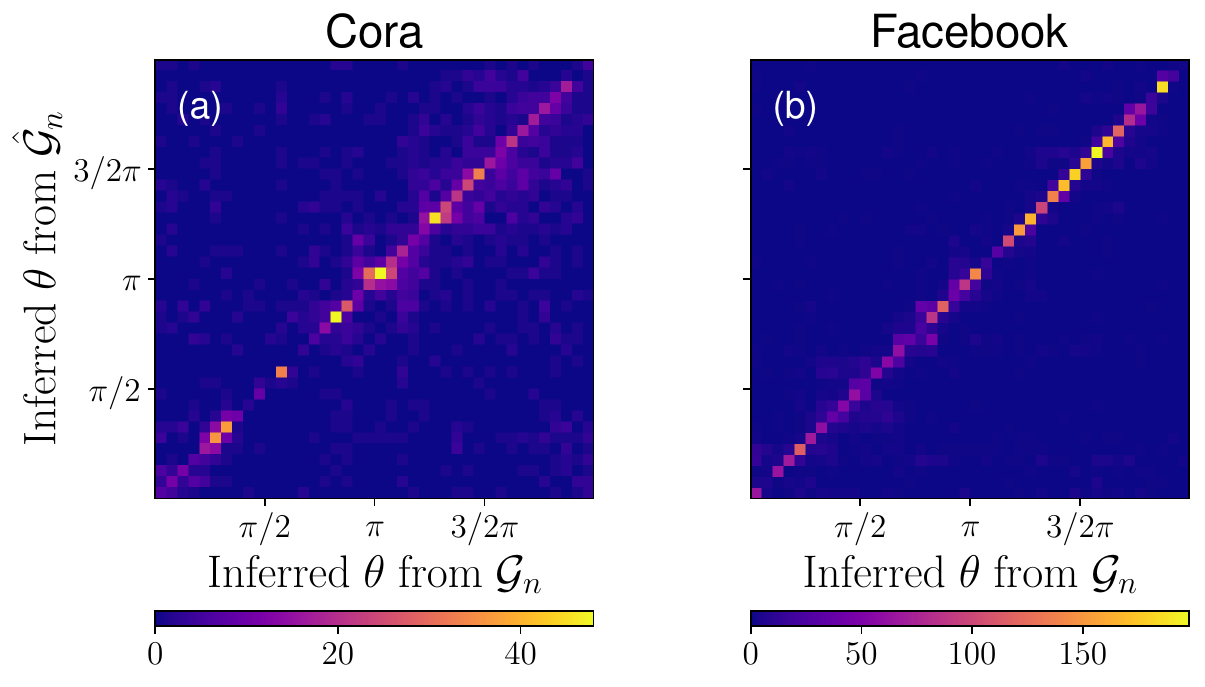}
	\caption{{\bf Detecting correlations between the spaces of nodes and features.} Heatmap of the angular coordinates of nodes inferred by Mercator from $\mathcal{G}_n$ (in the x-axis) and from $\mathcal{\hat{G}}_n$ (in the y-axis) for the Cora (a) and Facebook (b) datasets. A detailed description of these datasets is provided in Appendix~\ref{Appendix:B}. Color indicates the number of nodes in each pixel.}
	\label{fig:2}
\end{figure}
\begin{figure*}[t]
	\centering
	\includegraphics[width=\linewidth]{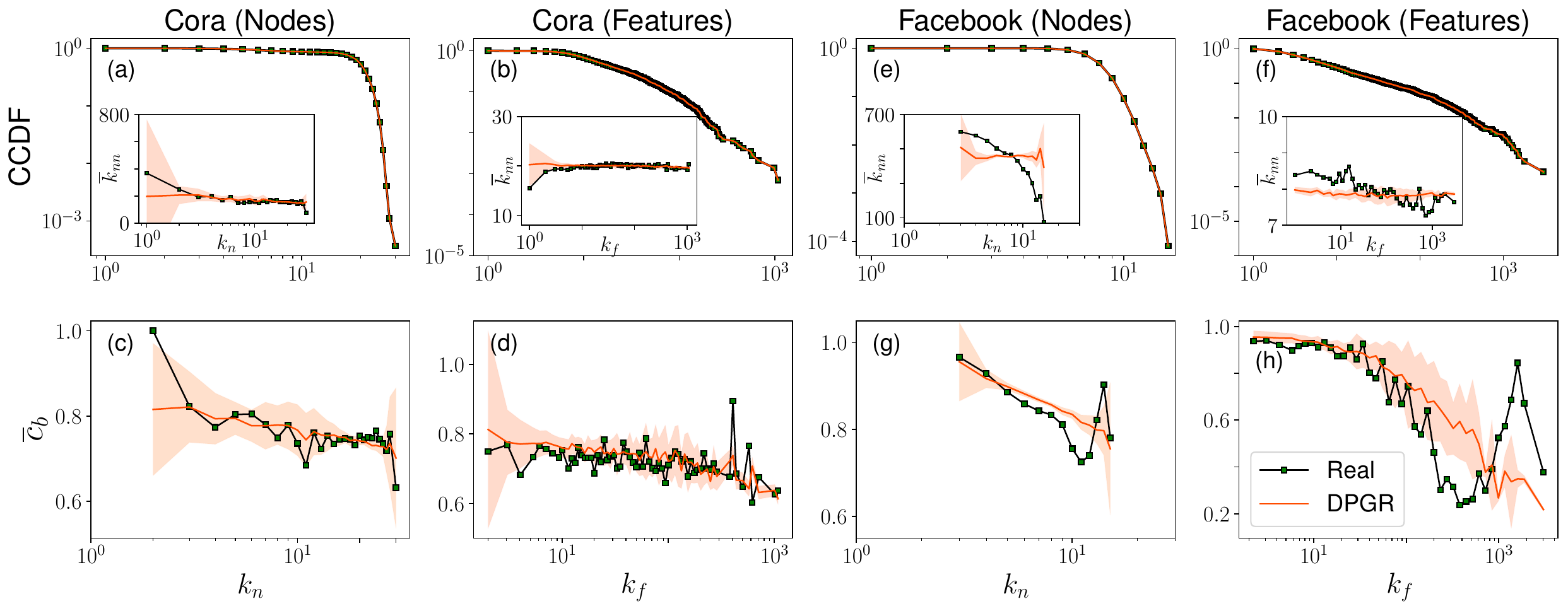}
	\caption{{\bf Topological properties of graph-structured data.} Topological properties of $\mathcal{G}_{n,f}$ for the Cora and Facebook datasets (symbols) and their synthetic counterparts generated by the bipartite-$\mathbb{S}^1$ model with the DPGR algorithm in Eq.~\eqref{DP:metropolis} (red solid lines). The top row (a,b,e,f) shows the complementary cumulative distribution functions of nodes and features degrees, whereas the insets in these plots show the average nearest neighbors degree functions $\bar{k}_{nn}$. These functions are defined, for features, as the average degree of nodes in $\mathcal{G}_{n,f}$, $k_n$, that are neighbors of features of degree $k_f$, and similarly for nodes. The bottom row (c,d,g,h) shows the bipartite clustering spectrum of nodes and features as a function of nodes and features degrees, respectively. The orange shaded area represents two-$\sigma$ intervals of the ensemble. Exponential binning is applied in the computation of $\bar{k}_{nn}$ and $\overline{c}_b$ for the features.}
	\label{fig:3}
\end{figure*}

\subsubsection{Detecting correlations between nodes and features}

As mentioned earlier, GCNs are effective when there is a correlation between the features of nodes and the underlying graph topology $\mathcal{G}_n$. Therefore, it is essential to identify this correlation in real-world datasets. To accomplish this, we define a new unipartite network between nodes, called $\mathcal{\hat{G}}_n$, where two nodes are connected if they share a significant number of features (for technical details, refer to Appendix \ref{Appendix:A}). It is important to note that the links in the networks $\mathcal{G}_n$ and $\mathcal{\hat{G}}_n$ are defined by different connection mechanisms so that, a priori, they could be unrelated. In order to measure any possible correlation between them, we assume that $\mathcal{\hat{G}}_n$ also follows the $\mathbb{S}^1$ model. Subsequently, the angular coordinates of nodes from $\mathcal{G}_n$ are inferred using Mercator~\cite{GarciaPerez2019}, and these coordinates are then employed as initial estimates to infer the angular coordinates of nodes from $\mathcal{\hat{G}}_n$, again using Mercator. 
The outcomes are depicted in Fig.~\ref{fig:2}, showcasing the results obtained from the Cora and Facebook datasets (additional information about these datasets can be found in Appendix \ref{Appendix:E}). The figure clearly illustrates a significant correlation between angular coordinates of nodes determined exclusively from topology and those determined exclusively from features. In contrast, randomized versions of $\mathcal{\hat{G}}_n$, which maintain the degree distribution and clustering coefficient, do not exhibit this correlation (see Appendix \ref{Appendix:A}). This empirical evidence strongly suggests that the similarity space of nodes and features is highly congruent.

\subsubsection{Geometric model of nodes and features. The bipartite-$\mathbb{S}^1/\mathbb{H}^2$ model}
Building upon this result, we propose our model for graph-structured data. The key aspect of our approach is to view the set of nodes and their features as a bipartite graph $\mathcal{G}_{n,f}$. In this representation, each node has a degree $k_n$ that indicates the number of distinct features it possesses, while each feature has a degree $k_f$ that represents the number of connected nodes. The top row of Fig.~\ref{fig:3} displays the complementary cumulative distribution function of node and feature degrees for the Cora and Facebook datasets. Across all the datasets we examined, we observed a consistent pattern characterized by a homogeneous distribution of node degrees and a heterogeneous distribution of feature degrees in the bipartite graph $\mathcal{G}_{n,f}$. The insets in these plots also reveal weak correlations between the degrees $k_n$ and $k_f$ of connected pairs.

\begin{figure}[th]
	\centering
	\includegraphics[width=\columnwidth]{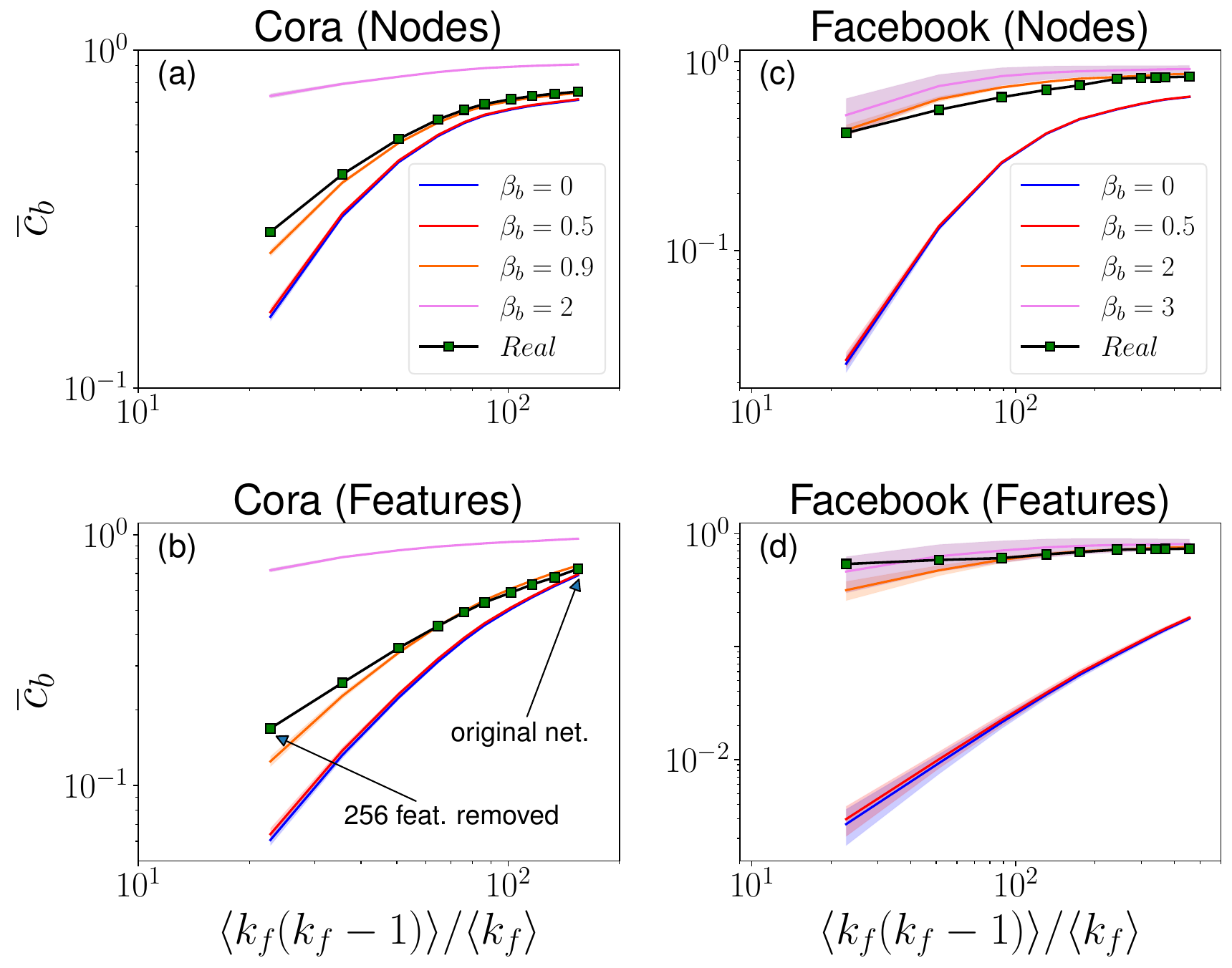}
	\caption{{\bf Behavior of bipartite clustering upon removal of hubs.} Bipartite clustering coefficient for the Cora and Facebook networks (symbols) and their surrogates generated by our model with different values of $\beta_b$ (solid lines). The plots show the bipartite clustering of the networks obtained by removal of a number of the highest degree features as a function of the corresponding fluctuations of features' degrees. In all plots, solid lines represent averages over 100 synthetic networks generated by our model.}
	\label{fig:4}
\end{figure}

Our objective is to develop a model for this bipartite graph that is correlated with the node network $\mathcal{G}_n$. To achieve this, we propose a geometric model called the bipartite-$\mathbb{S}^1$ model~\cite{serrano2012uncovering,kitsak2017latent}, where the similarity space is shared between $\mathcal{G}_n$ and $\mathcal{G}_{n,f}$, as suggested by the empirical correlation found in Fig.~\ref{fig:2}. In this model, each node is assigned two hidden variables $(\kappa_n, \theta_{n})$, where $\kappa_n$ represents its expected degree in the bipartite graph, and the angular coordinate corresponds to that of $\mathcal{G}_n$, i.e., $\theta_n = \theta$. Similarly, features are equipped with two hidden variables $(\kappa_f, \theta_f)$, indicating their expected degrees and angular positions in the common similarity space. The probability of a connection between a node and a feature with hidden degrees $\kappa_n$ and $\kappa_f$, separated by an angular distance $\Delta \theta$, is given by:
\begin{equation}
p_b(\kappa_n,\kappa_f,\Delta \theta)=\frac{1}{1+\chi^{\beta_b}} \;  \; \mbox{with}  \; \; \chi \equiv \frac{R \Delta \theta}{\mu_b \kappa_n \kappa_f},
\end{equation}
where $\mu_b=\frac{\beta_b}{2\pi \langle k_n \rangle} \sin{\frac{\pi}{\beta_b}}$ is a parameter determining the average degree of nodes $\langle k_n \rangle$ and features $\langle k_f \rangle=\frac{N_n}{N_f}\langle k_n \rangle$ (the sketch in Fig.~\ref{fig:1} illustrates the construction of the model). Similar to the $\mathbb{S}^1$ model, this choice ensures that the expected degrees of nodes and features with hidden degrees $\kappa_n$ and $\kappa_f$ are $\bar{k}_n(\kappa_n)=\kappa_n$ and $\bar{k}_f(\kappa_f)=\kappa_f$, respectively~\cite{serrano2012uncovering,kitsak2017latent}. The hidden variables of nodes and features can be generated from arbitrary distributions or fitted to replicate the topology of a real network of interest.

In the latter case, following the approach in~\cite{Starnini_2019}, it is also possible to define the ``microcanonical" ensemble of the model, by using a degree-preserving geometric randomization (DPGR) Metropolis-Hastings algorithm. This algorithm allows us to explore different values of $\beta_b$ while exactly preserving the degree sequences. Given a network and after assigning angular coordinates at random to all nodes and features, the algorithm randomly selects a pair of node-feature links $i_n-j_f$ and $l_n-m_f$, and swaps them (avoiding multiple connections) with a probability given by
\begin{equation}
p_{\text{swap}} = \min{\left[1,\left( \frac{\Delta \theta_{i_nj_f}\Delta \theta_{l_nm_f}}{\Delta \theta_{i_nm_f}\Delta \theta_{j_fl_n}}\right)^{\beta_b}\right]},
\label{DP:metropolis}
\end{equation}
where $\Delta \theta$ is the angular separation between the corresponding pair of nodes. This algorithm maximizes the likelihood that the network is generated by the bipartite-$\mathbb{S}^1$ model, while preserving the degree sequence and the set of angular coordinates. Notice that $\beta_b=0$ corresponds to the bipartite configuration model.

Similarly to the $\mathbb{S}^1$ model, the bipartite-$\mathbb{S}^1$ model can also be mapped to the hyperbolic plane. In this case, the mapping is given by
\begin{equation}
r_n=R^b_{\mathbb{H}^2}-2\ln{\frac{\kappa_n}{\kappa_{n,0}}}\; \; \mbox{and} \; \; r_f=R^b_{\mathbb{H}^2}-2\ln{\frac{\kappa_f}{\kappa_{f,0}}},
\end{equation}
where the radius of the hyperbolic disk is $R^b_{\mathbb{H}^2}=2\ln{\frac{2R}{\mu_b \kappa_{n,0}\kappa_{f,0}}}$, and $\kappa_{n,0}$ and $\kappa_{f,0}$ are the minimum values of the expected nodes and features degrees, respectively. With this mapping, the connection probability between a node and a feature depends only on the hyperbolic distance between them and takes the same form as in Eq.~\eqref{eq:pkkhyperbolic}.

\begin{table}[t]
\caption{Parameters of the bipartite network $\mathcal{G}_{n,f}$ for the analyzed datasets. Parameter $\beta$ for $\mathcal{G}_n$ is directly inferred by Mercator.}
\begin{center}
\begin{tabular}{|c|c|c|c|c|c|c|}
\hline
  & $N_n$ & $N_f$ & $\langle k_n \rangle$ & $\langle k_f \rangle$ & $\beta_b$ & $\beta$ \\
\hline
Cora & $2708$ & $1432$ & $18.174$ & $34.369$ & $0.9$ & $1.6$\\
Facebook & $12374$ & $3720$ & $7.542$ & $25.086$ & $2.0$ & $1.7$ \\
Citeseer & $3264$ & $3703$ & $31.745$ & $27.982$ & $0.9$ & $1.6$\\
Chameleon & $2277$ & $3132$ & $21.545$ & $15.663$ & $1.0$ & $1.6$\\
\hline
\end{tabular}
\end{center}
\label{table:1}
\end{table}

\subsubsection{Quantifying bipartite clustering in real datasets}

In the $\mathbb{S}^1$ model, the parameter $\beta$ governs the clustering coefficient and thus influences the relationship between the network topology and the underlying metric space. Similarly, the parameter $\beta_b$ accounts for the coupling between the bipartite graph $\mathcal{G}_{n,f}$ and the underlying metric space. As both $\mathcal{G}_n$ and $\mathcal{G}_{n,f}$ are defined on the same underlying metric space, the parameters $\beta$ and $\beta_b$ control the correlation between them. It is therefore important to measure the value of $\beta_b$ for a real dataset. To achieve this, we use the simplest possible extension of the clustering coefficient to bipartite networks, denoted as $\bar{c}_b$, as explained in the caption of Fig.~\ref{fig:1}. 

In bipartite networks, $\bar{c}_b$ is strongly influenced by the heterogeneity of the features' degree distribution and, for finite-sized networks, it can reach high values even in the configuration model (see Appendix \ref{Appendix:C}). Thus, measuring $\beta_b$ by adjusting $\bar{c}_b$ can be misleading and we took a different approach. 

\begin{figure}[t]
	\centering
	\includegraphics[width=0.9\columnwidth]{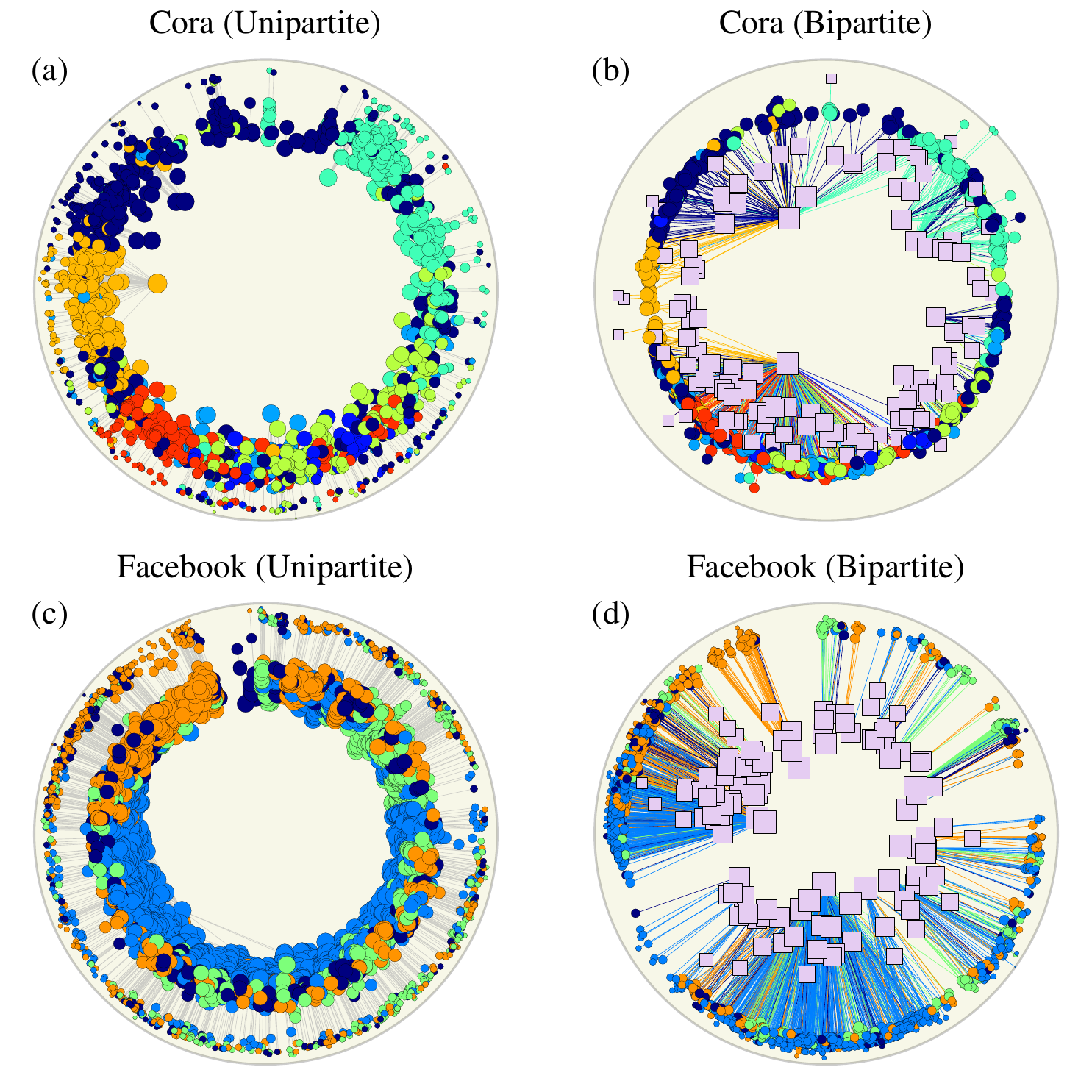}
	\caption{{\bf Hyperbolic representation of the Cora and Facebook datasets.} The left column shows hyperbolic embeddings of the unipartite network $\mathcal{G}_n$ and the right column the bipartite network $\mathcal{G}_{n,f}$ between nodes (colored circles) and features (pink squares). Nodes are colored by the labels given in the dataset. In both cases, we show edges with effective distance  $\chi<1$.Sizes of nodes and features are proportional to the logarithm of their degrees.}
	\label{fig:5}
\end{figure}

We sorted the degrees of features in decreasing order and removed  $2^l$ of the highest degree features from the original network, starting from the highest degree, where $l=0,1,2,\cdots$. After each removal, we measured the bipartite clustering coefficient $\bar{c}_b(l)$ of the remaining network and the fluctuations in features' degrees as $\langle k_f(k_f-1) \rangle / \langle k_f \rangle$ (see Appendix \ref{Appendix:C} for details). Fig.~\ref{fig:4} illustrates the behavior of the bipartite clustering for the Cora and Facebook datasets, considering values of $l$ up to $l_{\text{max}}=8$. We repeated this procedure for networks generated by our model with the DPGR algorithm Eq.~\eqref{DP:metropolis} and different values of $\beta_b$. Interestingly, for real networks, the bipartite clustering coefficient $\bar{c}_b(l)$ decreases slowly as hubs are removed. On the other hand, in the configuration model with $\beta_b=0$, $\bar{c}_b(l)$ decreases rapidly when the heterogeneity of the degree sequence is eliminated, even if the original network exhibits similar values to the real networks. As we increase the value of $\beta_b$, we observed that our model can accurately replicate the behavior of $\bar{c}_b(l)$, enabling us to estimate the values of $\beta_b$ in real networks. Beyond the practical estimation of parameter $\beta_b$, the slow decay of clustering when removing hubs provides strong empirical evidence that the bipartite network between nodes and features is governed by an underlying similarity metric space.

Table~\ref{table:1} presents the properties of the analyzed real networks and the inferred values of $\beta$ and $\beta_b$. Using these values, we generated network surrogates with the DPGR algorithm and compared their topological properties: degree distributions, degree-degree correlations, and bipartite clustering spectrum. Fig.~\ref{fig:3}  and Fig.~\ref{Fig5-SI} in Appendix ~\ref{Appendix:D} displays the ensemble average and the two-sigmas interval for all the measures. In all cases, the model accurately reproduces these properties. However, the model could be further improved by considering that nodes and features may not be uniformly distributed in the similarity space, but instead defining geometric communities, as discussed in~\cite{zuev2015emergence,garcia-perez:2018aa}.

Finally, we provide the hyperbolic representation of the Cora and Facebook datasets. First, we used Mercator to embed the unipartite network $\mathcal{G}_n$. The results are displayed in Fig.~\ref{fig:5} a and c, with nodes color-coded according to the labels provided in the dataset. As expected, we observe a high degree of congruence between these labels and the angular organization of nodes. Remarkably, this congruence serves as empirical evidence of the suitability of the $\mathbb{S}^1$ model for describing these networks. Notably, Mercator does not make use of these labels, making this congruence all the more significant. To represent the bipartite network $\mathcal{G}_{n,f}$, we maintain the angular coordinates of nodes from the embedding of $\mathcal{G}_n$ and adjust the hidden degrees of nodes and features in $\mathcal{G}_{n,f}$, $\kappa_n$ and $\kappa_f$, as well as $\beta_b$ to reproduce the degree distribution of nodes and features, as explained in detail in~\cite{GarciaPerez2019}. Subsequently, while keeping the coordinates of nodes fixed, we adjust the angular coordinates of each individual feature to maximize the likelihood that the feature is connected to its actual neighbors. The results are presented in Fig.~\ref{fig:5} b and d. In this case, nodes exhibit a homogeneous degree distribution, causing them to appear near the edge of the circle. Conversely, features exhibit a heterogeneous distribution, leading to a corresponding heterogeneous distribution of radial coordinates. Interestingly, features are closely connected to nodes within the same angular sectors. This suggests that by using this representation, it is possible to define specific sets of features associated with particular sets of nodes, thus defining bipartite communities. 

\section{Conclussions}

To summarize, our approach represents a paradigm shift in the description of complex graph-structured data. The crucial element in our framework is to view the relationships between nodes and features as a bipartite graph influenced by the same underlying similarity space that shapes the topology of the network between  nodes. We hypothesize that this shared similarity space, along with the strength of the coupling between networks $\mathcal{G}_n$ and $\mathcal{G}_{n,f}$ controlled by parameters $\beta$ and $\beta_b$, underlies the effectiveness of GCNs. If this conjecture holds true, our formalism could provide a crucial component in addressing the black box problem. On the short term, it is important to note that our results represents just an initial step towards an embedding tool for featured-enriched networks. The comprehensive embedding will emerge from the simultaneous maximization of the likelihood of both graphs, $\mathcal{G}_n$ and $\mathcal{G}_{n,f}$, a topic we will address in a forthcoming publication.

\section*{Acknowledgments}
We acknowledge support from: Grant TED2021-129791B-I00 funded by MCIN/AEI/10.13039/501100011033 and the  ``European Union NextGenerationEU/PRTR''; Grants PID2019-106290GB-C22 and PID2022-137505NB-C22 funded by MCIN/AEI/10.13039/501100011033; Generalitat de Catalunya grant number 2021SGR00856. M. B. acknowledges the ICREA Academia award, funded by the Generalitat de Catalunya.

\appendix

\section{Correlation between the inferred angular coordinates in $\mathcal{G}_n$ and $\mathcal{\hat{G}}_n$}\label{Appendix:A}

\begin{figure}[t!]
\centering
\includegraphics[width=\columnwidth]{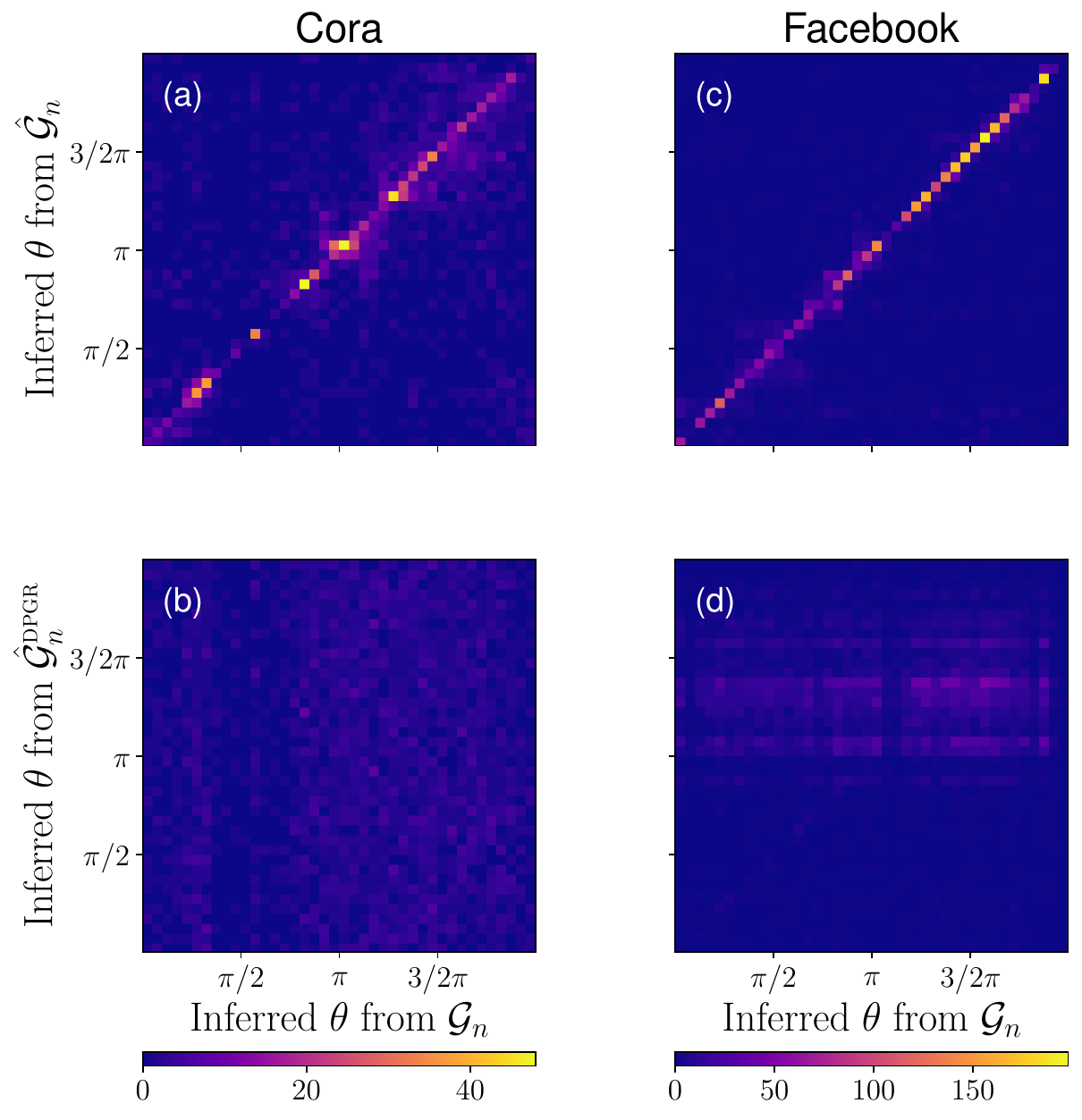}
\caption{Heatmap of the angular coordinates of nodes inferred by Mercator for the Cora and Facebook datasets. In all the plots, the X-axis shows the angular coordinates of nodes in $\mathcal{G}_n$. In the top row, the Y-axis corresponds to the angular coordinates of nodes in $\mathcal{\hat{G}}_n$, while in the bottom row, it represents the angular coordinates of a randomized version of $\mathcal{\hat{G}}_n$ called $\mathcal{\hat{G}}_n^{DPGR}$. For the Cora dataset, the significance level of $\alpha = 0.05$ in the disparity filtering method yields a backbone network consisting of $92\%$ of the nodes and $0.006\%$ of the links. For Facebook, using $\alpha = 0.03$ produces a backbone network with $72\%$ of the nodes and $0.003\%$ of the links.} \label{Fig1-SI}
\end{figure}
To evaluate the correlation between the features of nodes and the underlying graph topology $\mathcal{G}_n$, the unipartite network $\mathcal{\hat{G}}_n$ is extracted from the original bipartite network $\mathcal{G}_{n,f}$. The procedure entails the projection of $\mathcal{G}_{n,f}$ onto the set of nodes, resulting in a weighted unipartite network of nodes where the edge weights reflect the number of shared features between pairs of nodes~\cite{borgatti2014}. Typically, this yields a highly dense network with many spurious connections. Then, the disparity filtering~\cite{Serrano2009} is employed to capture the relevant connection backbone in this network.
 
The disparity filtering method normalizes the edge weights and determines the probability $\alpha_{ij}$ that an edge weight conforms to the null hypothesis, which assumes that the total weight of a given node is distributed uniformly at random among its neighbors. By applying a significance level $\alpha$, the links with $\alpha_{ij} < \alpha$ that reject the null hypothesis are deemed statistically significant and form the desired network $\mathcal{\hat{G}}_n$ along with their associated nodes.

The correlation between $\mathcal{G}_n$ and $\mathcal{\hat{G}}_n$ is assessed by assuming that $\mathcal{\hat{G}}_n$ follows the $\mathbb{S}^1$ model. The angular coordinates of nodes in $\mathcal{G}_n$ are determined using the Mercator embedding tool. Subsequently, these coordinates serve as initial estimates for inferring the angular coordinates of nodes in $\mathcal{\hat{G}}_n$ using Mercator once again. The process is performed for a total of 10 times, where in each iteration, the coordinates obtained from the previous step are utilized as the initial estimates. 

\begin{figure}[t!]
\centering
\includegraphics[width=\columnwidth]{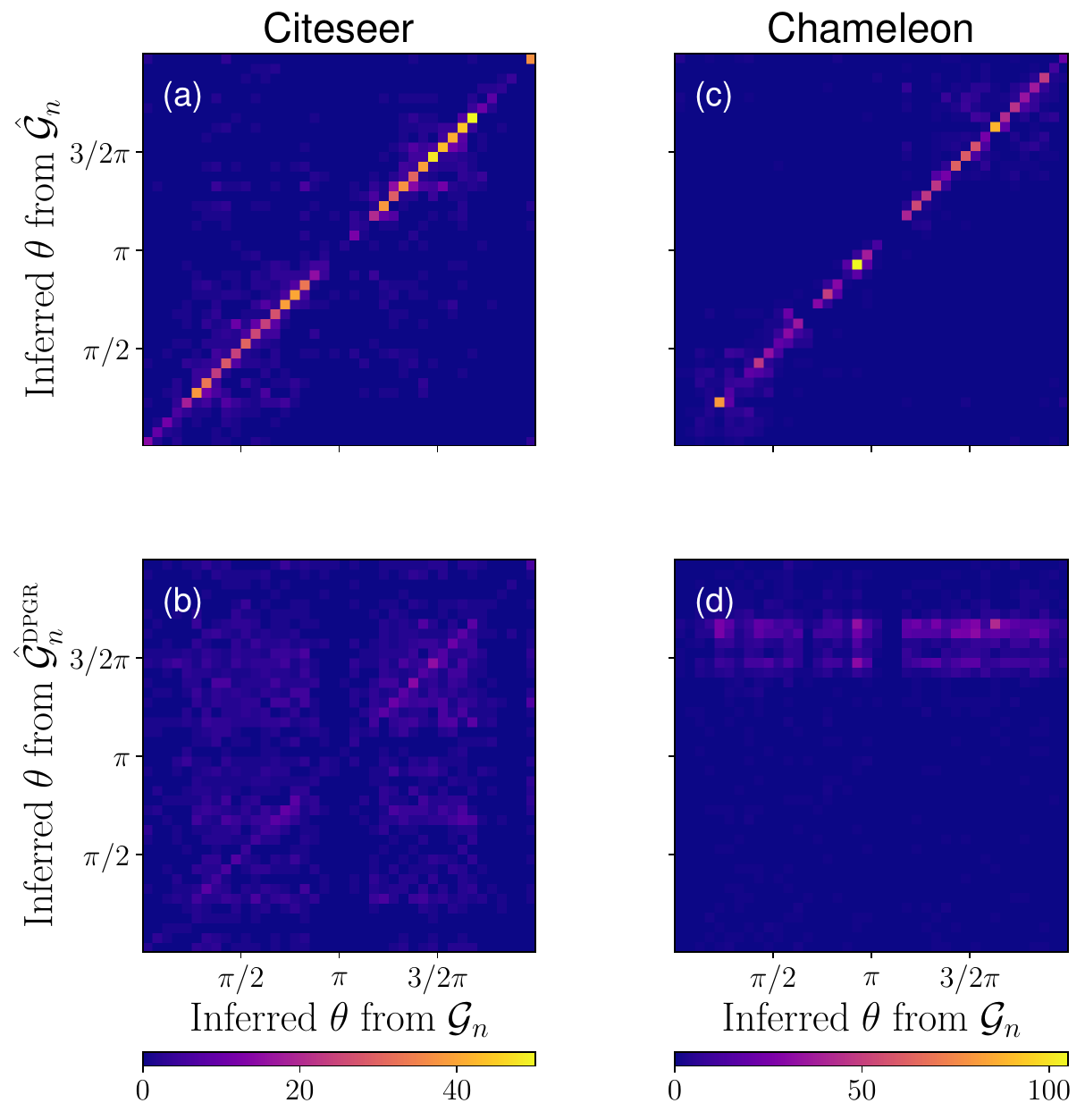}
\caption{Heatmap of the angular coordinates of nodes inferred by Mercator for the Citeseer and Chameleon dataset. In all the plots, the X-axis shows the angular coordinates of nodes in $\mathcal{G}_n$. In the top row, the Y-axis corresponds to the angular coordinates of nodes in $\mathcal{\hat{G}}_n$, while in the bottom row, it represents the angular coordinates of its randomized version $\mathcal{\hat{G}}_n^{DPGR}$. For the Citeseer dataset, applying a significance level of  $\alpha = 0.02$ results in a backbone network with $88\%$ of nodes and $0.003\%$ of edges. In the Chameleon dataset, setting $\alpha = 0.07$  generates a backbone network including $95\%$ of nodes and $0.023\%$ of links.}
\label{Fig2-SI} 
\end{figure}

In Fig.~\ref{Fig1-SI}, the top row clearly demonstrates a strong correlation between the angular coordinates of nodes of $\mathcal{G}_n$ and $\mathcal{\hat{G}}_n$ in the Cora and Facebook datasets. To rule out the possibility that this correlation is induced by the fact that we are using the angular coordinates of $\mathcal{G}_n$ as initial conditions to find the coordinates of $\mathcal{\hat{G}}_n$, we repeat the very same procedure with a randomized version of $\mathcal{\hat{G}}_n$, $\mathcal{\hat{G}}_n^{\text{DPGR}}$, generated using DPGR algorithm with the same $\beta$ that Mercator assigns to $\mathcal{\hat{G}}_n$. In this way, the randomized version has the same degree distributions and the same level of clustering as $\mathcal{\hat{G}}_n$. The bottom row of Fig.~\ref{Fig1-SI} shows no correlation between angular coordinates of $\mathcal{G}_n$ and $\mathcal{\hat{G}}_n^{\text{DPGR}}$, which proves that the correlation found in Fig.~\ref{Fig1-SI}~a and c is real and not an artifact of the method. Finally, Fig.~\ref{Fig2-SI} shows similar results for the Citeseer and Chameleon datasets. These experiments focuses on the angular coordinates of nodes within the giant components of networks.

\section{Dataset description}\label{Appendix:B}
\textbf{Cora\normalfont~\cite{Sen2008}}: It is a directed network of scientific publications, where an edge from $i_n$ to $j_n$ indicates that paper $i$ has cited paper $j$. Additionally, each paper is associated with a feature vector containing entries of either zero or one, which respectively show the absence or presence of specific words from a predefined dictionary. Therefore, a link between node $i_n$ and feature $m_{f}$ in the bipartite network signifies that the $m^{th}$ word from the dictionary has appeared in the paper $i$.

\textbf{Facebook\normalfont~\cite{musae2021}}: The network consists of Facebook pages categorized into four groups: politicians, governmental organizations, television shows, and companies. The links in the network represent mutual likes between these pages. Every page is assigned a node feature vector that is derived from its description, providing a summary of its purpose. These feature vectors indicate the presence or absence of specific words from a given bag of words. Accordingly, each node in the bipartite network is connected to the corresponding features associated with the words present in the page description. The degree distribution of nodes in the bipartite network is strongly bimodal. In this paper, in order to focus on one of the modes present in this distribution, we exclude nodes with more than 15 features. Subsequently, we remove these nodes from the unipartite network.

\textbf{Citeseer\normalfont~\cite{Sen2008}}: It is a directed citation network of papers where binary node features indicate whether specific words are present or absent in each paper. Consequently, in the unipartite network, each link between two papers signifies that one paper has cited the other. Similarly, in the bipartite network, a link between nodes and features denotes the inclusion of a specific word within the corresponding paper. 

\textbf{Chameleon\normalfont~\cite{musae2021}}: The network comprises Wikipedia articles centered around chameleons, where the connections represent mutual hyperlinks between the pages.  The binary feature vectors of the nodes imply the existence of informative nouns within the text of each Wikipedia article.

In this paper, we focus on simple graphs by removing self-loops and multiple links. We also convert directed networks into their undirected counterparts. Furthermore, we remove nodes and features with zero degrees, ensuring that only relevant and interconnected elements are considered.

\section{Bipartite clustering coefficient in the configuration model}\label{Appendix:C}

In a network of $N_n$ nodes and $N_f$ features generated by a bipartite soft configuration model, the connection probability between a node with expected degree $\kappa_n$ and a feature of expected degree $\kappa_f$ is given by
\begin{equation}
p_{\kappa_n,\kappa_f}=\frac{\kappa_n \kappa_f}{N_f \langle \kappa_f \rangle}=\frac{\kappa_n \kappa_f}{N_n \langle \kappa_n \rangle}.
\end{equation}

We define the bipartite clustering coefficient of a feature as the probability of two of its neighboring nodes being connected at least through a feature different from the one being analyzed. Using this definition, it is easy to see that the bipartite clustering coefficient of features for the soft configuration model is given by
\begin{equation}
\overline{c}_{b}^{~\text{features}} =  1 - \int \int \frac{\kappa_n \kappa_n' \rho_n(\kappa_n) \rho_n(\kappa_n')}{\langle \kappa_n \rangle ^ 2} e^{\frac{-\kappa_n \kappa_n' \langle \kappa_f^2 \rangle}{N_f \langle \kappa_f \rangle  ^2}} d\kappa_n d\kappa_n',\label{Eq1}
\end{equation}
where we have used that the probability that a feature of expected degree $\kappa_f$ is connected to a node of expected degree $\kappa_n$ is $\rho(\kappa_n | \kappa_f)= \kappa_n \rho_n(\kappa_n)/\langle \kappa_n \rangle$ and where $\rho_n(\kappa_n)$ is the distribution of expected degrees of nodes. Analogously, the bipartite clustering coefficient of nodes is given by
\begin{equation}
\overline{c}_{b}^{~\text{nodes}} =  1 - \int \int \frac{\kappa_f \kappa_f' \rho_f(\kappa_f) \rho_f(\kappa_f')}{\langle \kappa_f \rangle ^ 2} e^{\frac{-\kappa_f \kappa_f' \langle \kappa_n^2 \rangle}{N_n \langle \kappa_n \rangle  ^2}} d\kappa_f d\kappa_f'. \label{Eq2}
\end{equation}

In the soft configuration model $\langle \kappa_f \rangle =  \langle k_f \rangle$ and $\langle \kappa_{f}^2 \rangle =  \langle k_f (k_f -1) \rangle $, and $\langle \kappa_n \rangle =  \langle k_n \rangle$ and $\langle \kappa_{n}^2 \rangle =  \langle k_n (k_n -1) \rangle $ where $k_f$ and $k_n$ are the actual degrees of nodes and features, respectively. 

Empirical measures show that quite generally the bipartite graphs $\mathcal{G}_{n,f}$ are characterized by homogeneous node degree distributions. Thus, we assume that the distribution of hidden nodes' degrees is distributed by a Dirac delta function, that is, $\rho_n(\kappa_n) = \delta (\kappa_n  - \langle \kappa_n \rangle)$. In this case, Eq.~\eqref{Eq1} become
\begin{equation}
\overline{c}_{b}^{~\text{features}} =  1 -  e^{\frac{-\langle \kappa_n \rangle \langle \kappa_f^2 \rangle}{N_n \langle \kappa_f \rangle}} \label{Eq3}
\end{equation} 

\begin{figure}[t!]
	\centering
	\includegraphics[width=\columnwidth]{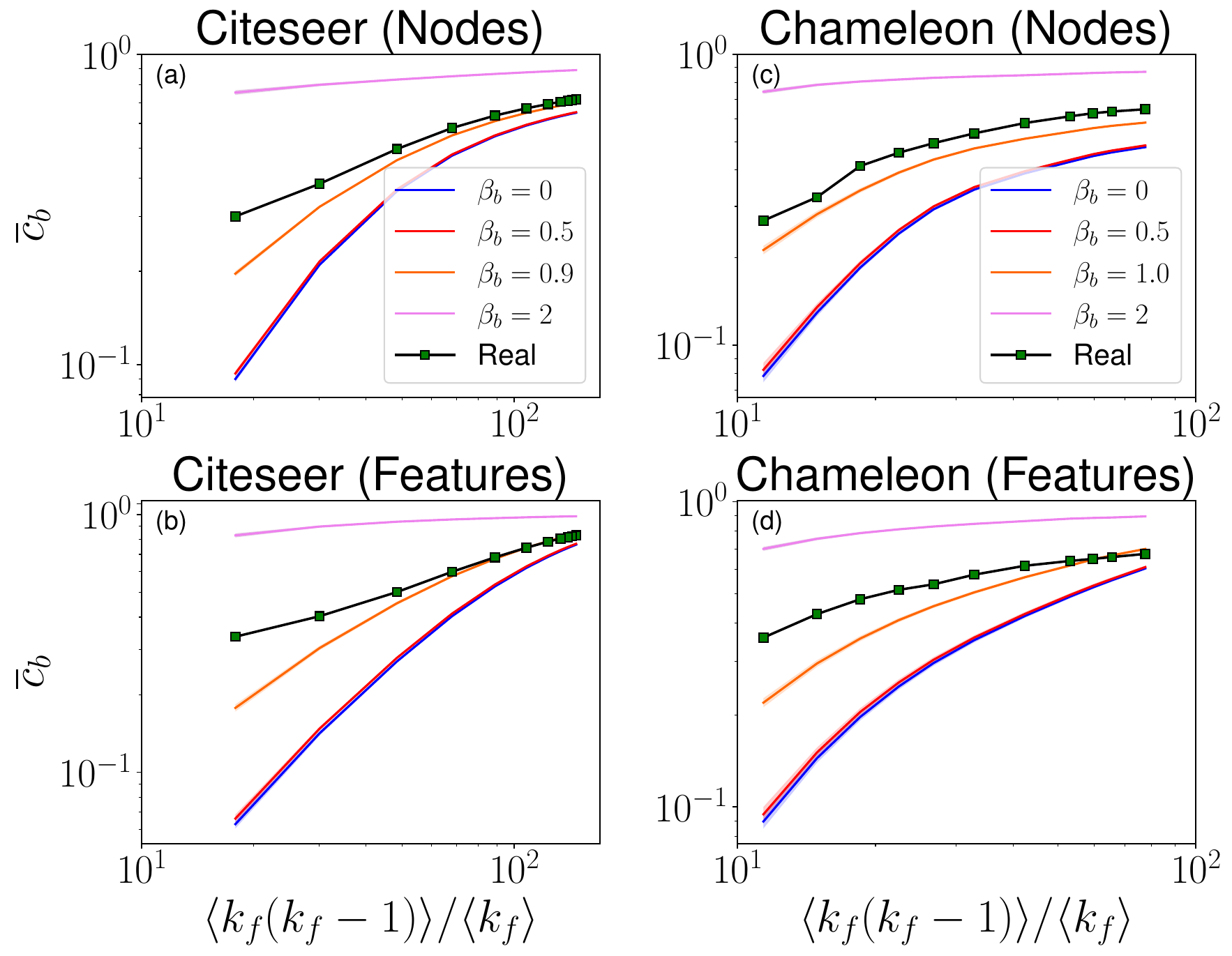}
	\caption{Bipartite clustering coefficient for the Citeseer and Chameleon networks (symbols) and their surrogates generated by our model for different values of $\beta_{b}$ (solid lines). The plots show the bipartite clustering of the networks obtained by removal of a number of the highest degree features as a function of the corresponding fluctuations of features' degrees. The solid lines represent the average bipartite clustering over 100 synthetic networks generated by our model.}
	\label{Fig3-SI}
\end{figure}

The bipartite clustering coefficient for nodes in Eq.~\eqref{Eq2} cannot be, in general, further simplified unless we specify $\rho_f(\kappa_f)$. However, for not very heterogeneous distributions of features' degrees, and in the thermodynamic limit it reads as
\begin{equation}
\overline{c}_{b}^{~\text{nodes}} = \frac{1}{N_n} \frac{\langle k_f (k_f - 1) \rangle ^2}{\langle k_f \rangle ^2}  \label{Eq4}
\end{equation}
In all cases, the bipartite clustering coefficient of nodes increases with the heterogeneity of the distribution of features' degrees. By introducing the variable $x \equiv \frac{\langle k_f(k_f - 1)\rangle}{\langle k_f \rangle}$, Eqs.~\eqref{Eq3} and~\eqref{Eq4} can be rewritten in terms of $x$ as
\begin{equation}
\overline{c}_{b}^{~\text{features}} = 1- e ^ {\langle k_n \rangle x / N_n}  \label{Eq5}
\end{equation}
\begin{equation}
\overline{c}_{b}^{~\text{nodes}} = \frac{x^2}{N_n}, \label{Eq6}
\end{equation}
which highlights that bipartite clustering is strongly influenced by the heterogeneity of the features' degree distribution, and for finite-sized networks it can be very large due to the high value of $x$. 

At the light of these results, to detect significant clustering in real datasets, we propose a sequential approach in which $2^l$, $l= 0, 1, 2, ...$,  of the features  with the highest degrees are consecutively removed from the original real-world network. The bipartite clustering coefficient of the resulting network $\overline{c}_b(l)$ is then plotted as a function of the fluctuations in features' degrees, expressed by $ \frac{\langle k_f(k_f - 1)\rangle}{\langle k_f \rangle}$. The experimental results for the Citeseer and Chameleon datasets in Fig.~\ref{Fig3-SI} illustrate that in real-world networks, as hubs are progressively removed, $\overline{c}_b(l)$ exhibits a slow decrease. Conversely, in bipartite configuration networks generated by our model with $\beta_b = 0$ in the DPGR algorithm, $\overline{c}_b(l)$ shows a rapid decline as the  heterogeneity of the features' degree is reduced. By increasing the value of $\beta_b$, our model effectively replicates the behavior of $\overline{c}_b(l)$, enabling us to estimate the bipartite clustering coefficient in real-world networks.

\newpage

\onecolumngrid

\section{Topological properties of unipartite networks $\mathcal{G}_{n}$}\label{Appendix:E}
\begin{figure*}[h]
	\centering
	\includegraphics[width=\textwidth]{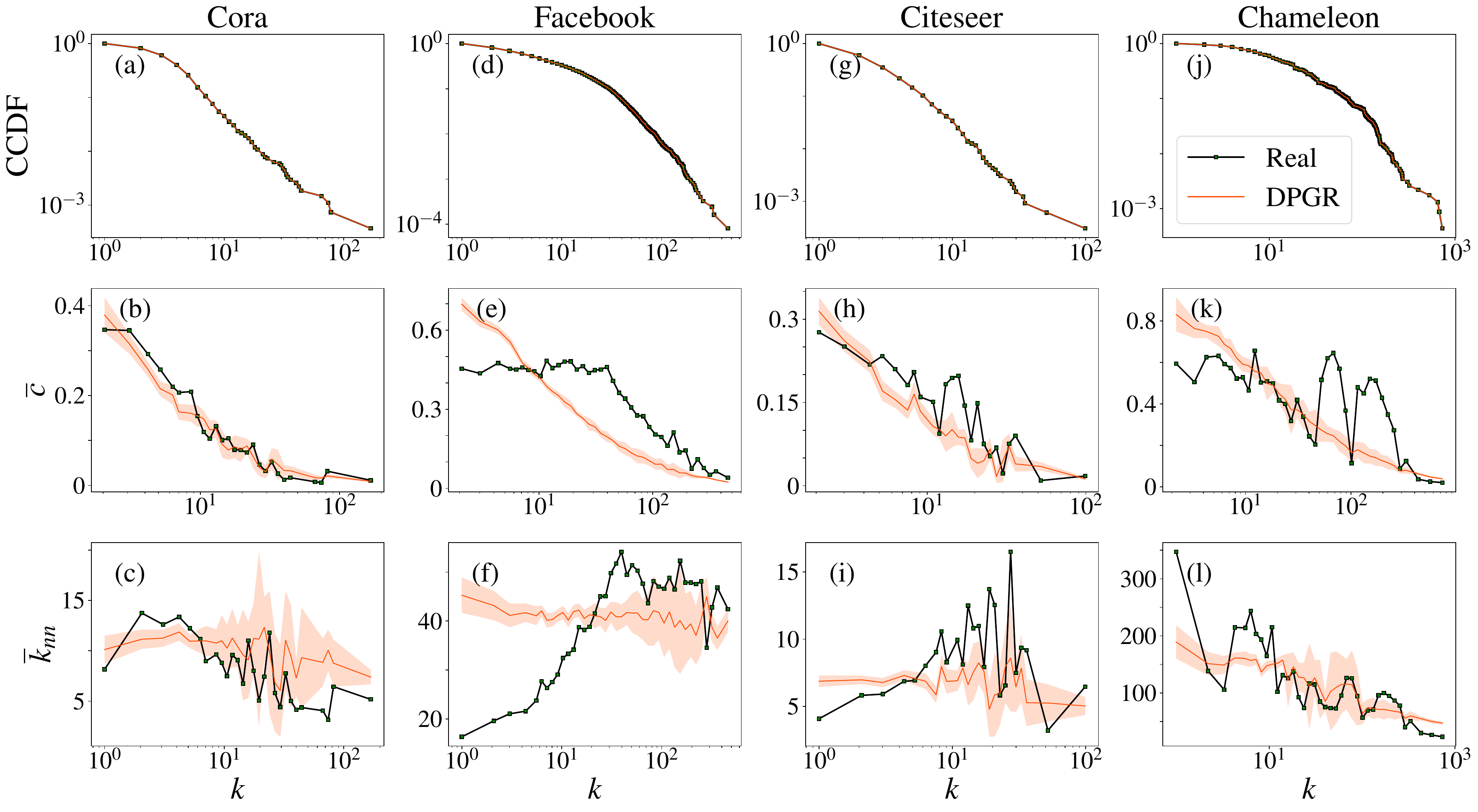}
	\caption{Topological properties of $\mathcal{G}_{n}$ for all datasets (symbols) and their synthetic counterparts generated by the $\mathbb{S}^1$ model using DPGR method (red solid lines). The top row (a-j) shows the complementary cumulative distribution functions of nodes. The middle row (b-k) represents the average nearest neighbors degree functions, and the bottom row (c-l) shows the clustering spectrum as a function of node degrees. Exponential binning is applied in the computation of $k_{nn}$ and $\overline{c}$. The orange shaded area represents two-$\sigma$ intervals around the mean for 100 realizations of the model.}
	\label{Fig4-SI}
\end{figure*}

\newpage

\section{Topological properties of bipartite networks $\mathcal{G}_{n,f}$}\label{Appendix:D}

\begin{figure*}[h]
\centering
\includegraphics[width=\textwidth]{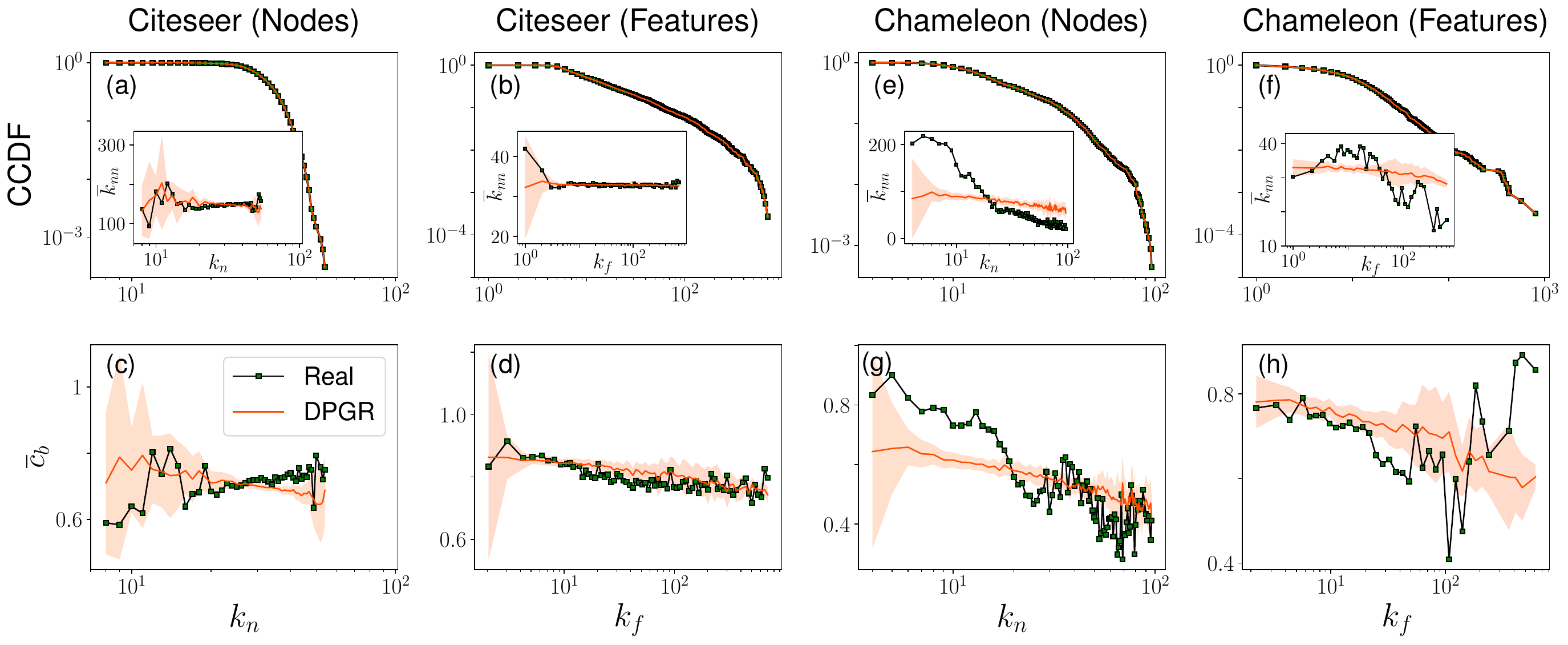}
\caption{Topological properties of $\mathcal{G}_{n,f}$ for the Citeseer and Chameleon datasets (symbols) and their synthetic counterparts generated by the bipartite-$\mathbb{S}^1$ model (red solid lines). The top row (a-f) shows the complementary cumulative distribution functions of nodes and features degrees, whereas the insets in these plots show the average nearest neighbors degree functions. The bottom row (c-h) shows the bipartite clustering spectrum as a function of nodes and features degrees. Exponential binning is applied in the calculation of $k_{nn}$ and $\overline{c}_b$ for the features. The orange shaded area represents two-$\sigma$ intervals around the mean for 100 realizations of the model.}\label{Fig5-SI}
\end{figure*}

\appendix

\twocolumngrid 
\newpage
\bibliography{geometry3}

\end{document}